# Fabrication and Characterization of High-Purity Germanium Detectors with Amorphous Germanium Contacts

X.-H. Meng[a], G.-J. Wang[a], M.-D. Wagner[a], H. Mei[a], W.-Z. Wei[a], J. Liu[a], G. Yang[a], D.-M. Mei[a],[*]

[a]*Department of Physics, The University of South Dakota,*
*414 E. Clark Street, Vermillion, South Dakota 57069, USA*
*E-mail:* Dongming.Mei@usd.edu

ABSTRACT: Large, high-purity, germanium (HPGe) detectors are needed for neutrinoless double-beta decay and dark matter experiments. Currently, large (> 4 inches in diameter) HPGe crystals can be grown at the University of South Dakota (USD). We verify that the quality of the grown crystals is sufficient for use in large detectors by fabricating and characterizing smaller HPGe detectors made from those crystals. We report the results from eight detectors fabricated over six months using crystals grown at USD. Amorphous germanium (a-Ge) contacts are used for blocking both electrons and holes. Two types of geometry were used to fabricate HPGe detectors. As a result, the fabrication process of small planar detectors at USD is discussed in great detail. The impact of the procedure and geometry on the detector performance was analyzed for eight detectors. We characterized the detectors by measuring the leakage current, capacitance, and energy resolution at 662 keV with a Cs-137 source. Four detectors show good performance, which indicates that crystals grown at USD are suitable for making HPGe detectors.

---

[*] Corresponding author.

# Contents

## 1. Introduction

Cosmogenic produced isotopes can limit the sensitivity for large-scale germanium-based (Ge-based) experiments in the search for dark matter and detection of neutrinoless double-beta decay [1-4]. For example, $^{3}$H, $^{49}$V, $^{56}$Fe, and $^{65}$Zn, produced by cosmogenic activation when the Ge detectors are fabricated on the surface, are main sources of background events in the MAJORANA DEMONSTRATOR and EDELWEISS in the low energy region of interest [5-6]. Similarly, $^{60}$Co and $^{68}$Ge can be the sources of background events in the higher energy region for the detection of neutrinoless double-beta decay [7]. An effective way to reduce the production of cosmogenic isotopes in Ge is to grow Ge crystals and fabricate detectors underground at the site where the experiments will take place.

Since the successful development of lithium-drifted Ge detectors introduced the significant use of semiconductor crystals for direct detection and spectroscopy of gamma ray in the 1960s [8-13], high-purity Ge (HPGe) detectors gradually became a standard technology to achieve spectroscopy or imaging of gamma rays by providing the best compromise between energy resolution and efficiency for high resolution gamma-ray spectroscopy [14-17]. A small bandgap energy of Ge (~0.7 eV) creates a large number of electron–hole pairs during interaction with gamma rays, which provides good energy resolution. Commercially available large HPGe crystals (up to 10 cm in diameter) enhance the probability of total absorption of an incoming gamma ray in the crystal leading to a high detection efficiency [18-19]. Currently, HPGe crystals are not only the best choice of material for gamma-ray spectroscopy but also a well-accepted technology for rare event physics in the search for dark matter [12-13, 20-21] and neutrinoless double-beta decay [22-29]. Therefore, HPGe detectors have been used in several research projects, including CoGeNT[30-31], SuperCDMS [32-34], EDELWEISS [35-37], GERDA [38-40], MAJORANA DEMONSTRATOR [41-42], CDEX[21, 43-44], focused on detecting dark matter or neutrinoless double-beta decay.

In order to make HPGe crystal growth and detector fabrication in an underground laboratory possible, the University of South Dakota (USD) has developed a research and development program (R&D) under the support of the Department of Energy and the state of South Dakota. After seven years R&D, large size (~5 inches in diameter) HPGe crystals have now been grown at USD [45-46].

One kind of simple detector used solely for spectroscopy of gamma-ray radiation is made of a single piece of HPGe crystal on two opposite surfaces on which two electrical contact layers are fabricated. These electrical contacts are used for the application of bias voltage and signal readout and must be able to block hole and electron injection enough to reduce electronic noise and achieve low leakage current [47-48]. A very reliable and well-established process employed in industry to manufacture such contacts utilizes boron (B) implantation to form an electron-blocking contact and lithium (Li) diffusion to form a thick and robust hole-blocking contact [26, 49-50]. This technology has been applied in a wide range of applications from basic science to commercial activities [18]. However, due to the thickness and significant diffusion of Li-diffused contacts at room temperature [50-51], this technology presents a challenge in forming finely segmented detectors. These are complex detectors used to measure energy and determine the positions of radiation interaction events in the entire detector for applications requiring imaging or particle tracking in addition to spectroscopy. The minimum thickness of the Li-diffused contact is about 1 mm, which creates undesirable effects for application in underground experiments such as neutrinoless double-beta decay and dark matter search [52].

An alternative technology developed at Lawrence Berkeley National Laboratory (LBNL) employs a amorphous-semiconductor (a-Ge or a-Si) contact, which is capable of providing finely segmented contacts on HPGe detectors with both electron and hole blocking properties [53-57]. This technique can replace the commercialized technology of Li-diffused and B-implanted contacts. In addition, the fabrication processes for detectors using amorphous semiconductor contacts is much simpler than employing Li-diffused and B-implanted contacts [57-60]. The amorphous semiconductor electrical contact technology is generating more and more interest and attention in both basic science and industry [61-63]. Benefiting from the pioneers at LBNL who have explored the amorphous-semiconductor contact technology, USD has developed a program to study Ge detector performance with a-Ge contacts fabricated from USD-grown crystals.

This paper describes the manufacturing process employed at USD including shaping a home-grown large HPGe crystal into a small planar detector, manual lapping and chemical etching, sputtering of a-Ge contact, and depositing a thin aluminum (Al) layer by using an electron-beam evaporator. We also study the planar detector performance so that we can explore the properties of the HPGe crystals grown at USD and provide feedback to our crystal-growth group for improving techinques for the growth of high-quality crystals. In our group, the HPGe crystals were grown through the Czochralski method using the zone-refined ingots produced at USD from commercial raw materials [64-67]. The growth process and the characterization method were described in several papers from our group [45, 68-70]. In this paper, we will focus on the fabrication of the detectors.

## 2. Experimental methods

To investigate the quality of the HPGe crystals grown at USD for use as detector-grade crystals and to have an accurate determination of their impurity concentration, a few small planar detectors have been fabricated at USD. All crystals converted into detectors are p-type with an impurity concentration ranging from ~$5 \times 10^9$ to ~$5 \times 10^{10}$/cm$^3$. The concentration is measured using the Hall Effect for the top and the bottom of the crystal. Note that the Hall Effect measurements possess uncertainty caused by the size of the contacts and the control of temperature. Therefore, the impurity concentration measured with the Hall Effect provides a reference point at which the crystal is justified to be pure enough to make a detector. Two different geometries of planar detectors were designed as shown in Figure 1. The thickness of their grooves and wings are fixed at 1.5 mm and 2 mm, respectively. The fabrication process is the same for the detectors with different geometries.

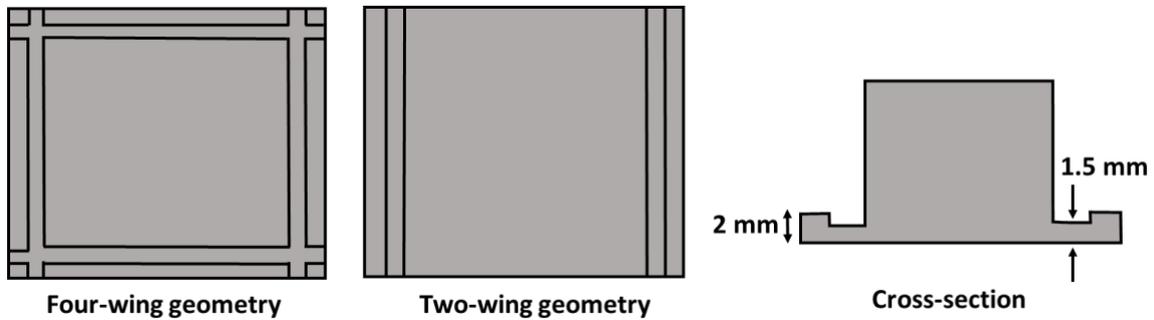

**Figure 1.** Schematic of detector geometries for top view and cross-section.

### 2.1 Crystal cutting

The cutting process starts with a large HPGe crystal, such as the one displayed in figure 2a. This crystal was grown using the Czochralski method in a hydrogen atmosphere. First, a segment of the desired purity in the portion between S1 and S2 as depicted in Figure 2a is sliced from the crystal using a diamond wire saw. The slice shown in Figure 2b comes from the portion between S1 and S2. This large segment was further cut into several smaller pieces based on design dimensions. During this process, a graphite plate is necessary for the cuts that pass through the HPGe crystal to prevent the diamond saw blade from cutting into a metal plate of stainless steel. For mounting the small crystal, a hot plate was used to heat sticky wax to hold the graphite plate between a stainless-steel plate and the HPGe crystal. Figure 2c shows several well-cut small pieces after the cutting process. Once the pieces were cut, the hot plate was used again to warm up wax to release the graphite plate and the small crystals. The same process was used to mount the small piece of crystal on the graphite plate onto a new stainless-steel plate for mechanically grinding and cutting in order to make a planar detector. A 2 mm thick blade was used to grind the wings and grooves of the planar detectors with an automatic feed setting of 0.5 mm/min typically used to produce a clean cut. A cutting fluid is continually sprayed onto the blade and crystal during the process to keep them cool. A well-cut four-wing detector is shown in Figure 2d. This crystal-graphite-steel stack was then heated to remove the crystal which was immediately cleaned off to remove wax, using wipes.

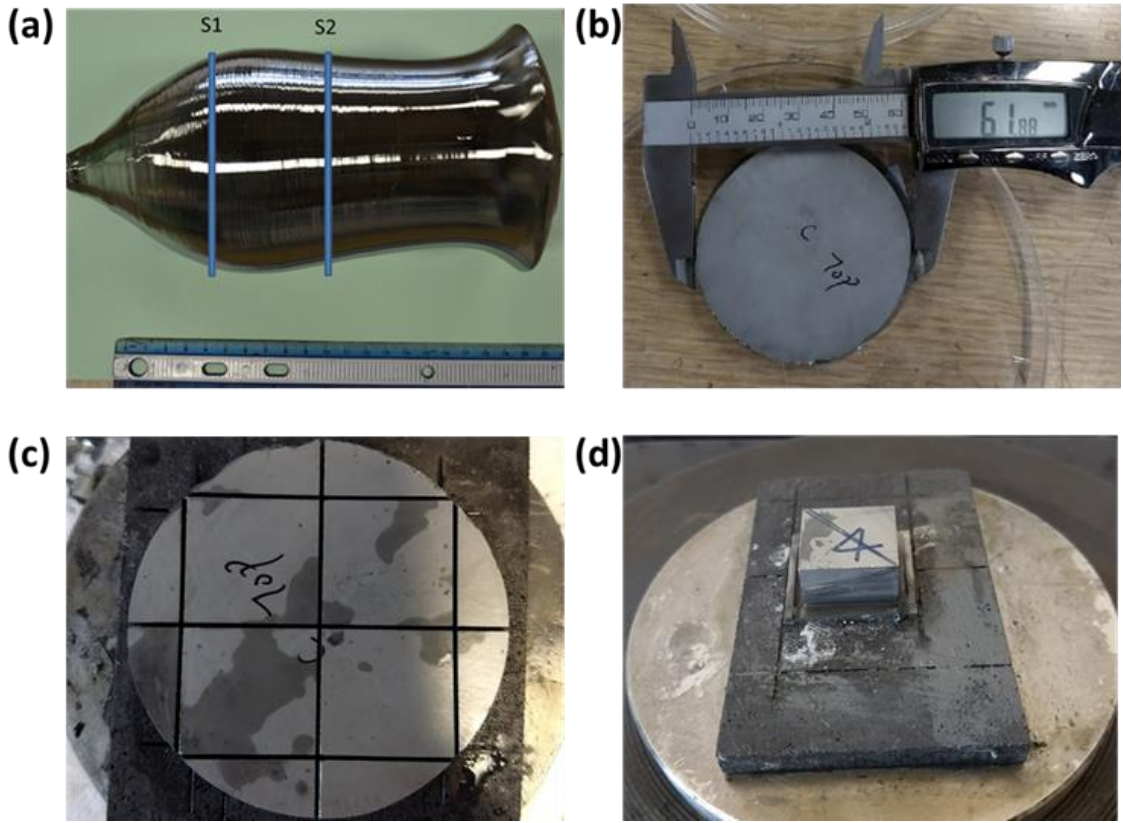

**Figure 2.** As described in the text (a) A large HPGe crystal grown at USD; (b) slice of crystal (a) cut between S1 and S2; (c) Small square shape crystals after cutting; (d) Four-wing planar crystal.

**2.2 Crystal lapping and chemical polish etching**

**2.2.1 Manual lapping process**

To remove any blade marks left by the cutting operation, each of the exposed surfaces of the cut crystal was then lapped. Crystal lapping includes coarse and fine lapping. Coarse lapping can quickly remove the chips and scratches from the top and bottom surfaces of the crystal. If both surfaces are smooth and flat without any visible chips and scratches, a coarse lapping is not necessary and only a fine lapping is required. Before lapping, a well-cut crystal with the desired shape must be cleaned with trichloroethylene (TCE) to completely remove the wax from the entire surface of crystal.

Lapping a crystal requires a big glass plate coated with a slurry composed of a teaspoon of grit lapping powder mixed with the distillate (DI) water. Micro abrasives 17.5 µm SiC powder is used for coarse lapping and 9.5 µm $Al_2O_3$ powder is used for fine lapping. During the coarse lapping, a gentle downward pressure can be applied to the crystal being manipulated in figure-eight or circular motion. This process can be repeated to completely lap away any chips at the edges of the crystal. Both top and bottom surfaces must be lapped until the entire surface has a uniform texture. Then the crystal is thoroughly rinsed with DI water. The fine lapping process can be used on both top and bottom surfaces to achieve a fine, uniform texture of the surfaces which helps further the chemical polishing process. Fine lapping can be done directly on the glass plate covered with the fine slurry, or on a fabric pad, which is put on the glass plate and then is covered with the fine slurry, as shown in Figure 3a. During this process, a small circular motion can be used with no downward pressure added so that a scratch-free surface can be finally obtained. Figure 3b displays a crystal thoroughly rinsed with DI water and dried using nitrogen gas ($N_2$) after both coarse and fine lapping processes are accomplished. The entire surfaces must be very clean without any water residue or stain.

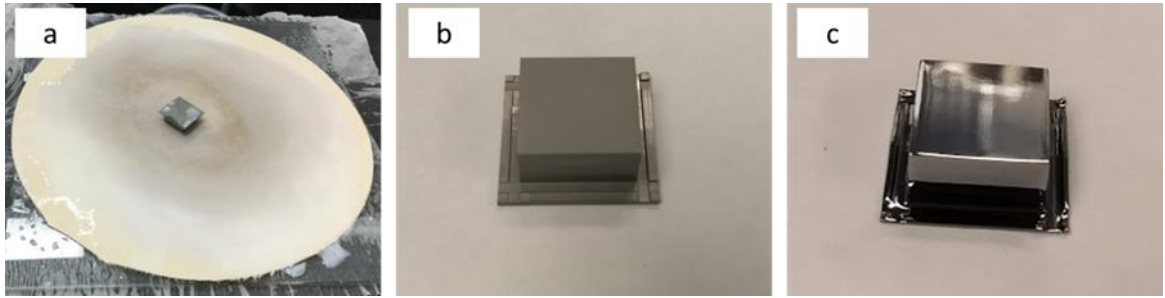

**Figure 3.** (a) Manually lapping process on a fabric pad covering a glass plate. (b) A fine-lapped crystal after water clean and $N_2$ dry. (c) A crystal after chemical polish.

**2.2.2 Chemical polishing process**

After the lapping process, a chemical polishing process is necessary to achieve a proper surface on the entire crystal. Figure 3c shows a crystal with a shiny surface obtained after the chemical polishing process. This process requires a strong acid etchant, a mixture of concentrated nitric acid ($HNO_3$) and hydrofluoric acid (HF) at a volumetric ratio of 4:1. Such a strong etchant can be held in only an acid-resistant beaker such as one made of Teflon. We complete the whole chemical polishing process in a fume hood while wearing personal protective equipment (PPE) and two layers of gloves as shown in the insert picture of Figure 4a. In Figure 4a, there are three Teflon beakers for the chemical polish process. One beaker holds the strong acid etchant. The other two contain DI water for thoroughly rinsing the crystal after etching. The first etching process, long term etching, takes around 3 minutes right after the lapping and removes any pits and invisible scratches from the lapping. During the etching process, the crystal is placed directly in the beaker containing enough etchant to cover the whole crystal and is continuously and rapidly moved around in the etchant at the bottom of beaker by rocking the beaker in a circular motion. The crystal also must be flipped several times during the etching period of about 3 minutes. When the etching time is over, the etched crystal is quickly taken out of the etchant using long tongue tweezers and is immediately soaked in DI water to quench the etching process. It should be consecutively rinsed for several ten second periods in the two beakers containing DI water. Then high purity $N_2$ gas is used to completely dry the shiny crystal. If there are chips or cracks at the crystal's edges that have not been etched smooth, one must lap these edges away and repeat the etching process. If the surface of the entire crystal is not shiny enough, or has partially cloudy areas, one can just repeat the long-term etching. The etchant can be reused for this additional etching as long as it has not become too warm and is not producing fumes. At last, the well-etched crystal should have a smooth and shiny mirror-like surface.

To remove any invisible flaws from the etched crystal, one short term etching process of about 30 seconds is necessary, since the etched crystal may touch an absorbent paper during the inspection of its surface after the long-term etching. Freshly-prepared etchant and DI water are

– 5 –

required for the short-term etching. Then, the crystal is held and manipulated in the fresh etchant using long tongue etching tweezers for the whole etching process lasting 30 s, followed by two separate rinsings in DI water, subjected to the $N_2$ drying process, and then being directly loaded into the sputtering jig as shown in Figure 4c. During the etching and subsequent processing steps, the crystal was held by the tweezers and was not allowed to touch any other surfaces. Figure 4b shows the $N_2$ drying process with an adsorbent paper under the crystal with no contact between the crystal and paper.

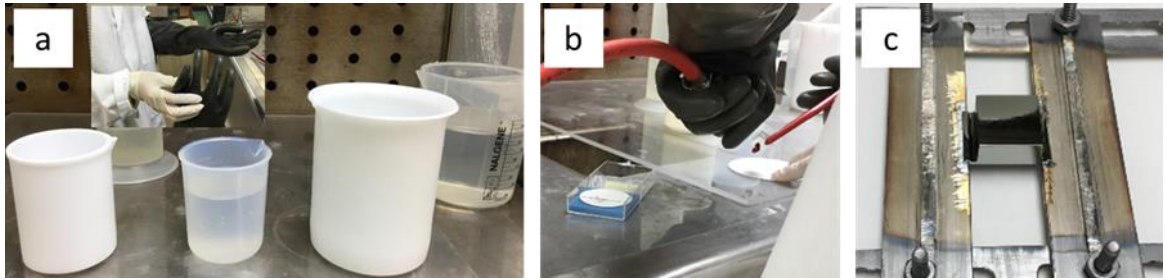

**Figure 4.** (a) Set up for etching process. Insert shows double layer gloves for etching. (b) $N_2$ dry process after short time etching with no contact between the crystal and paper. (c) Crystal was loaded into sputtering jig right after etching process.

### 2.3 Deposition of contact layers

### 2.3.1 Sputtering coat of a-Ge contact

A-Ge electrical contact was fabricated by using sputtering system (Perkin-Elmer) model 2400 as shown in Figure 5a. After a successful etching process, the crystal was directly loaded into the jig of the sputtering machine and then immediately put into the sputtering system. It was carefully surrounded by an aluminum (Al) foil mask to avoid back-sputtering of Ge atoms onto the bottom of the crystal. Figure 5b and Figure 5c display how the crystal was covered by Al mask for the sputtering process. A high vacuum, which is usually below $4 \times 10^{-6}$ Torr, is obtained by cryopumping the chamber for about 4 hours. A gas mixture of argon and hydrogen (Ar-7% $H_2$) was used for sputtering the a-Ge contact. A typical set of conditions are 14 mTorr chamber pressure measured by a 275 convectron gauge calibrated for $N_2$, 100 W forward power, and 0 W reflected power. The top and the four-side surfaces of the crystal were sputtered first. Pre-sputtering was taken for 5 minutes on the shutter and then sputtering deposition occurred for 15 minutes while 10 °C cooling water was recycling in the instrument. Another 15 minutes is required for cooling after the deposition. Then the crystal is ready to flip over and sputter the bottom surface with the same process as for the top surface. The crystal can be removed from the chamber after it has been cooled. It is then directly moved into the E-beam machine for Al evaporation. The sputter target used is 8 inches in diameter and composed of 99.999% purity Ge obtained from our crystal-growth group.

We measured the thickness of the a-Ge contact on the top surface of the detector and found it to be about 620 nm. The coat on the sides was uneven with a thickness of 350 nm at the top and gradually decreasing to 250 nm. This variation in thickness could come from shadows due to the static sputtering geometry. The Alpha-Step Profiler (KLA Tencor) was used to measure the thickness of the deposited layer. The profiler works by running a needle from a-Ge coating region to the uncoated area and thus measures the thickness of the deposited layer.

– 6 –

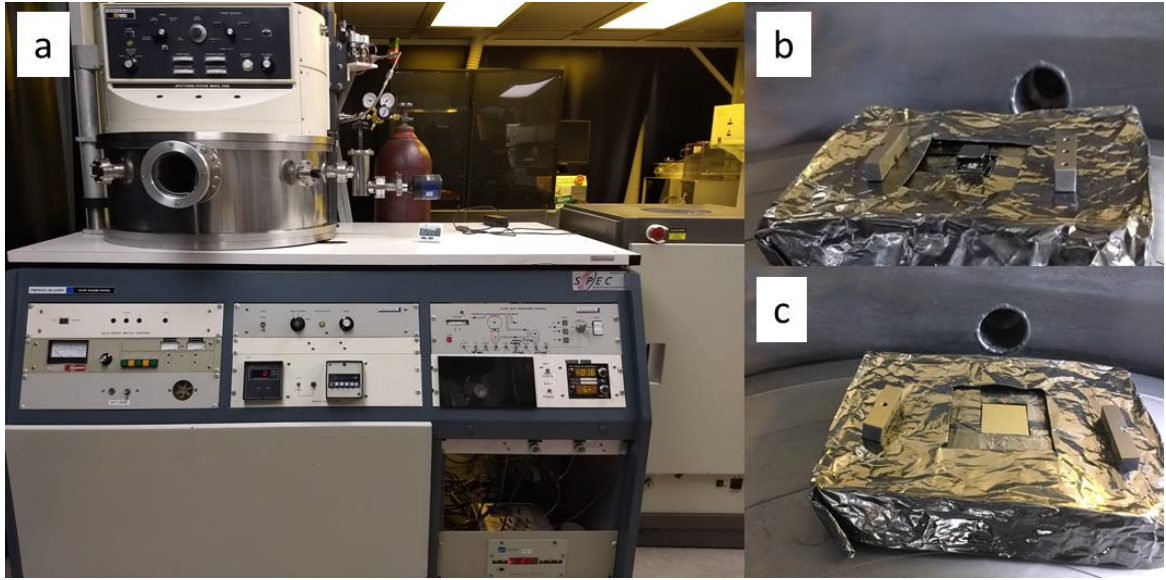

**Figure 5**. (a) Sputtering instrument at USD. (b) a-Ge deposition on the top of crystal. (c) a-Ge deposition on the bottom of crystal.

**2.3.2 Deposition of Al layer**

Figure 6a shows the instrument for the Al layer deposition as the readout electrode layer. A sample holder in Figure 6b was redesigned to hold HPGe crystals while avoiding any handling scratches. A high vacuum level of $10^{-6}$ mbar was required for the evaporation of the Al layer. The specific set of conditions for our instrument were 4.89 kV high voltage, 0.2 ~ 0.4 nm/s of deposit rate, with a thickness of 100 nm. After the evaporation on one surface was done, the crystal has to undergo a 45-minute cooling process before it is taken from the chamber and flipped over to coat the Al layer onto the another surface. There is no primary order for Al layer coating on the top and four side surfaces or bottom surface. Note that the crystal can be held only by the crystal handle and must not touch any other surfaces during the flipping process.

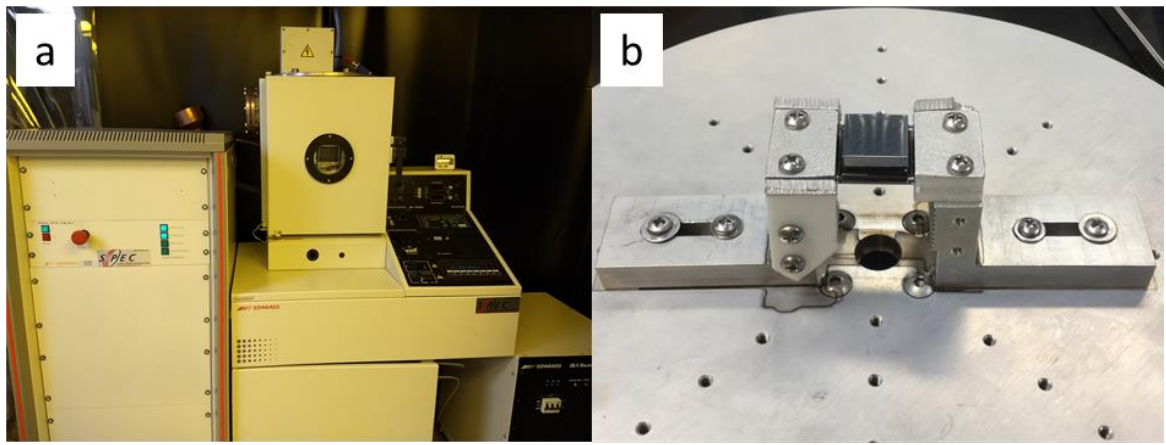

**Figure 6.** (a) Edwards EB3 Electron Beam Evaporator at USD. (b) Designed sample holder for HPGe detector.

## 2.4 Final etching of detector

The final step in the HPGe detector fabrication process is to remove the Al layer from the surface of the four sides so that the electric field lines mainly run from top to bottom. A small injection leakage current from only two contacts (top and bottom) can be achieved to increase energy spectroscopic signals. Acid resistant tape is employed to cover both the top and bottom surfaces as shown in Figure 7a. A cotton swab is used to provide a small amount of pressure on the tape to avoid the formation of air bubbles, which may cause the etchant to leak into the space between the tape and Al layer coated surface. The protected detector is then submerged into HF dip (1%) solution for around 2 minutes while long tongue tweezers are used to agitate the detector in the etchant. Such agitation contributes to the removal of the gas bubbles from the exposed surfaces and boosts the etching process of Al layer from the side surfaces. When the etching time has elapsed, the detector is immediately taken out from the etchant and quickly put in the DI water to quench the etching process and then is rinsed for several ten-second intervals with DI water. Afterwards it is thoroughly dried by blowing pure $N_2$ gas over the entire surface. Figure 7b shows the cross-section of a fabricated HPGe detector. The a-Ge contact covers all surfaces of the crystal. The Al layer is coated on the cryistal's top and bottom surfaces only.

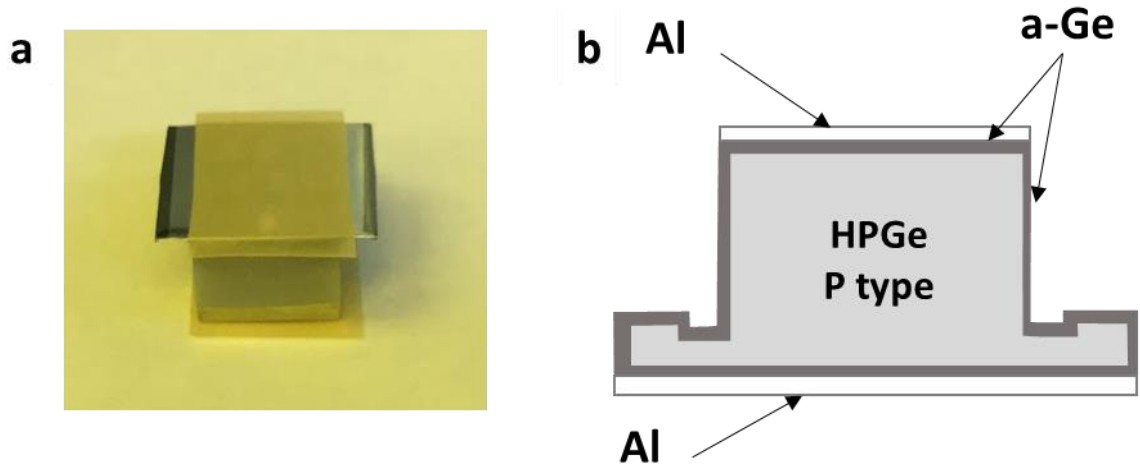

**Figure 7**. (a) A detector covered by acid resistant tap. (b) Cross-section of a fabricated detector.

## 2.5 Detector characterization

After each detector was fabricated, it was immediately loaded onto the sample stage in a test cryostat, as displayed in Figure 8a. As described in our recent paper [71], such a cryostat was specially designed and built at our collaboration lab (LBNL) so that the detector and variable temperature stage are enclosed by an infrared shield. The temperature of the sample stage can be controlled in the range from 79 K to around 300 K by a thermal controller. Liquid $N_2$ was used to cool the detector so that the capacitance and the leakage current could be measured at 79K. The measurement electronics for the characterization of the detectors includes a multimeter connected to a transimpedance amplifier for leakage current measurement and signal processing electronics for the readout of the signals. The signal readout electronics consisted of an AC-coupled charge-sensitive preamplifier followed by a commercial analog pulse-shaping amplifier. Such a signal readout is able to take the spectral characterization of the detector and the measurement of detector capacitance as a function of the applied detector voltage (C-$V_a$ characteristic). Figure 8b shows the external connection of electronics for the characterization of the detectors.

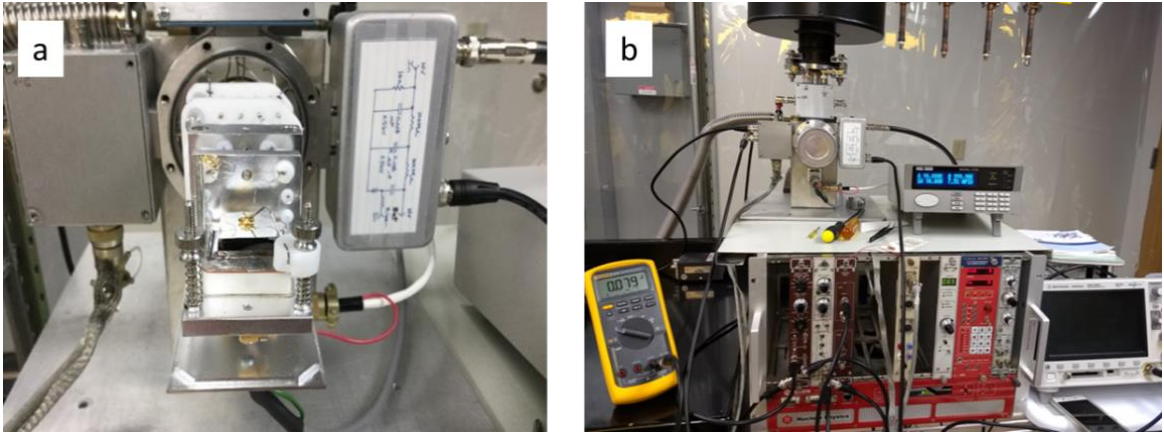

**Figure 8.** (a) A test cryostat loaded a detector. (b) External set up for detector characterization.

Each detector was first measured for leakage current as a function of applied voltage ( I-V curve ) to determine whether the detector was of sufficient quality to hold the high applied voltage, enabling it to reach the full depletion voltage. Then the C-$V_{ap}$ characteristic was measured to determine the full depletion voltage and the impurity concentration of the crystal. We applied a negative voltage to the bottom contact so that depletion began at the top contact of the detector. The energy spectrum was taken with ORTEC MCA. It was calibrated using a 662 keV peak and one of the X-ray peaks, 31.85 keV from a Cs-137 source. The pulser peak was taken to display the electronics noise level of the test system. The full width of half maximum (FWHM) of 662 keV and pulser peak were analysized by MCA software.

## 3. Results and discussion

Table 1. A summary of geometry and performance for all planar Ge detectors made with crystals grown at USD in the past half year. Detector investigation was completed at liquid $N_2$ temperature.

| Detector | geometry | Dimensions (L*W*D cm³) | Calculated Impurity (cm⁻³) | Leakage Current @Full Depletion Voltage (pA) | Full Depletion Volatge (V) | Capacitance (pF) | Energy Resolution (FWHM) | Noise (FWHM) |
|---|---|---|---|---|---|---|---|---|
| USD-L01 | four wings | 18.60*12.40*5.40 | 3.945E10 | 1 | 650 | 5.09 | 1.40 keV @ 662 keV | 0.97 |
| USD-L02 | four wings | 17.09*17.56*9.26 | 3.945E10 | Abnormal leakage current behavior (a fine line crossing the edge of detector) | | | | |
| USD-L03 | two wings | 24.70*18.70*6.00 | 7.73E9 | Unable to be fully depleted since the contact cannot hold detector bias voltage greater than 370V. | | | 4.39 keV@ 662 keV | 1.28 |
| USD-L04 | four wings | 21.07*19.72*10.70 | | Hold voltage up to 3700 v. Too thick to deplete at applied bias lower than 3000 V | | | 1.67 keV @ 662 keV | 1.2 |
| USD-L05 | two wings | 21.16*20.26*10.70 | | Unable to be fully depleted since the contact cannot hold detector bias voltage greater than 630V. | | | | |
| USD-L06 | two wings | 20.05*20.07*8.48 | 2.954E10 | 1 | 1200 | 6.52 | 2.22 keV @ 662 keV | 1.67 |
| USD-L07 | four wings | 21.53*20.77*8.48 | 2.461E10 | 1 | 1000 | 5.25 | 1.59 keV @ 662 keV | 1.19 |
| USD-L08 | four wings | 20.21*19.93*8.48 | 1.969E10 | 2 | 800 | 4.79 | 1.38 keV @ 662 keV | 1.03 |

Note: FWHM of both 662 keV and pulser peak, respectively displaying energy resolution and electronics noise level. The impurity concentration was calculated based on I-V and C-V measurement.

In this section, we present results of detector measurement and analyze some possible reasons for the failure of a detector. Table 1 provides information about geometry and detector performance for the eight planar detectors made from crystals grown at USD. Some detectors were reprocessed and tested many times to improve the properties of contact layers [17].

**3.1 Sputtering jig impact on detector performance**

Using a sputtering jig designed for 4-wing crystals to sputter a two-wing crystal may cause some problems in detector performance. Eight small planar detectors were fabricated using the same process, five with four-wing geometry and three with two wings. USD-L06, a two-wing detectors was successfully fabricated and displayed satisfactory detector performance. USD-L03 USD-L05, also two wing detectors, could not hold high voltage. However, USD-L04 with four wings, fabricated from the same HPGe crystal as USD-L05, could hold high voltage, up to 3700 V, while still not reaching full depletion voltage. This can be understood through a relation between the depth of the depletion versus the applied bias voltage for a given impurity level as described below: $d = \sqrt{2\varepsilon_{Ge}\varepsilon_0 v_b/eN_{|A-D|}}$, where $d$ represents the thickness of the depleted region, $\varepsilon_{Ge}$ is the relative permittivity of Ge, $\varepsilon_0$ is the permittivity of free space, $v_b$ is the applied bias voltage, $e$ stands for the electron charge in coulombs, and $N_{|A-D|}$ is the net impurity level in the detector. This relation indicates that the detector, USD-L04, was too thick to be fully depleted at 3700 V. To fully delplete this detector with a thickness of 1.07 cm for a given impurity level of ~$4\times10^{10}$/cm$^3$, the required bias voltage would exceed 4000 volts, which is beyond the 3000 V applied voltage capability of our test bench. However, the 3700 V holding-voltage of USD-L04 displays that the contact layers were successfully fabricated on the detector.

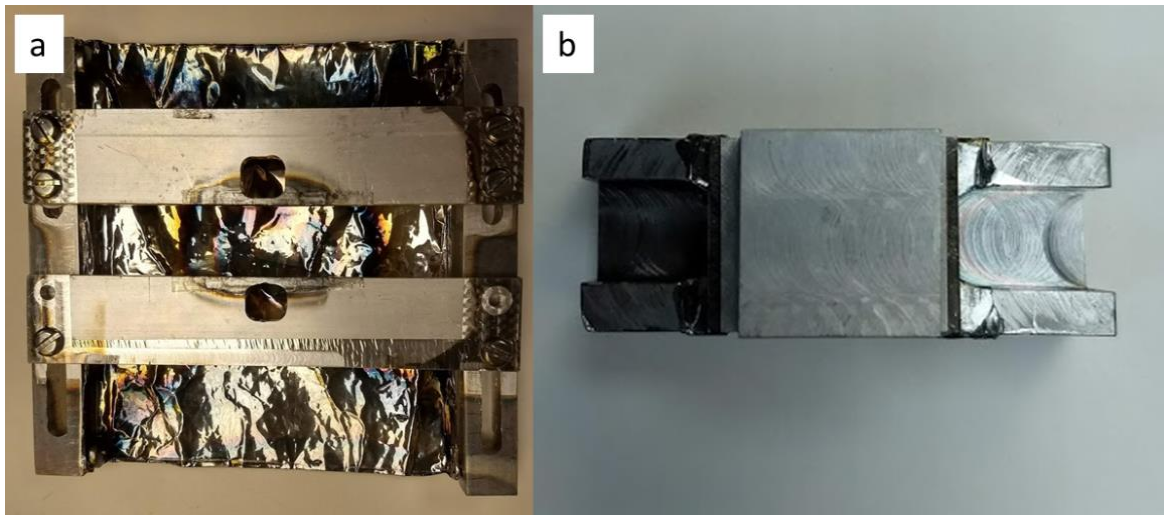

**Figure 9.** (a) Sputtering jigs for four-wing detectors. (b) Sputtering jigs for two-wing detectors.

The two-wing geometry design can significantly reduce cutting time. However, two-wing detectors require a two-wing sputtering jig for a-Ge deposition to avoid back-sputtering of Ge atoms onto the bottom surface of crystal. Such a jig is specific to the size of the detector and the two or four wing design. When we used the four-wing sputtering jig on two-wing detectors, this poorly fitted jig allowed many a-Ge atoms to back-sputter on the lower surface of crystal. Such back-sputtered spots can cause the two-wing detector's failure to hold high voltage.

Four-wing detectors take more cutting time than two-wing detectors. However, the jig required for four-wing detectors is adjustable to a wide range of detector sizes. The four wings of the detector help to prevent the back-sputtering of Ge atoms on the bottom surface, which makes the sputtering coat of four-wing a-Ge contact much easier. Figure 9 shows two different sputtering jigs for four-wing geometry design (Figure 9a) and two-wing geometry design (Figure 9b).

Although a two-wing detector, USD-L06, displayed normal detector behavior, its performance was not as good as other four-wing detectors. Figure 10a shows the I-V curve of USD-L06. The leakage current was 0 pA when a voltage of up to 1000 V was applied. It started to increase slightly while the bias voltage was higher than 1200 V. However, the leakage current increased quickly as bias voltage was raised from 1300 V to 1600 V. It reached 100 pA at 1600 V. The full depletion voltage of USD-L06 was around 1200 V as shown in Figure 10b. Once a detector reaches its full depletion voltage, its capacitance becomes constant with increasing voltage. A detailed analysis has been reported recently in one of our papers[71]. As mentioned in the experimental part, a negative bias was applied to the bottom of detector so that the detector starts the depletion from the top surface gradually reaching the bottom surface. The leakage current of USD-L06 began to increase rapidly when the applied bias was higher than 1200 V which means that the depletion has just reached the bottom surface of the detector. Such a rapid increase in the leakage current is related to electron injection from the bottom contact. The a-Ge contacts of USD-L06, were finished using four-wing jigs. The un-winged sides of the bottom surface likely suffered a back-sputtered a-Ge when a-Ge deposition was underway on the top surface. Such a back-sputtered a-Ge area may also have caused the failure of detector fabrication for USD-L03 and USD-L05 or largely affected the detector performance. To make a successful two-wing detector, one must use a jig constructed specifically for sputtering two-wing detectors.

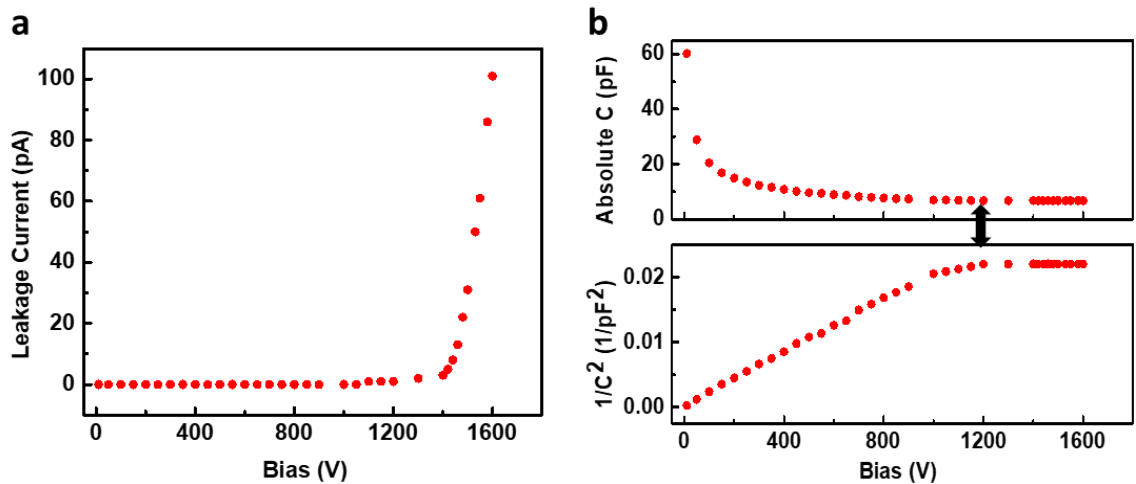

Figure 10. (a) I-V curve and (b) C-V curve for two-wing detector USD-L06.

## 3.2 Defect Impact

The USD-L02 detector displayed abnormal behavior during the leakage current measurement. Its leakage current did not show a natural rise and fall when a bias voltage was applied. Eventually, this detector could not hold a high bias voltage. This weird phenomenon may be related to a very fine linear defect that crossed the edge and extended from the top surface to the side surface as indicated in Figure 11a by the white arrow. We reprocessed this detector beginning with manual lapping followed by the chemical etching. This reprocess was repeated three times. Each time, no visible uneven features appeared around the defective area after manually lapping. However, such a linear defect appeared again after 3 minutes of long term etching. An extended longer-term etching, around 7 minutes, was employed to remove this defect. The result was that the longer etching caused a worse defect. A microscope was used to look into the defect area after manual lapping and chemical etching. Figure 11b showed the microscopic image of the defect area. No other nonuniform features appeared on the rough surface since it was lapped using micro-abrasive powder. After chemical etching, a uniform linear defect appeared as shown in Figure 11c. This defect may have been caused during the cutting process by excessive feeding speed, which was 2 mm/min, or by environmental vibration since a powerful air-compressor was very close to the slender, diamond cutting-saw. A slow feeding speed and anti-vibration condition would help to avoid such cutting damage. Such a defect may also be caused by a crystallographic defect. Overall, once a small crack appears on the surface of a crystal after chemical etching, one has to keep lapping till all damage is completely removed.

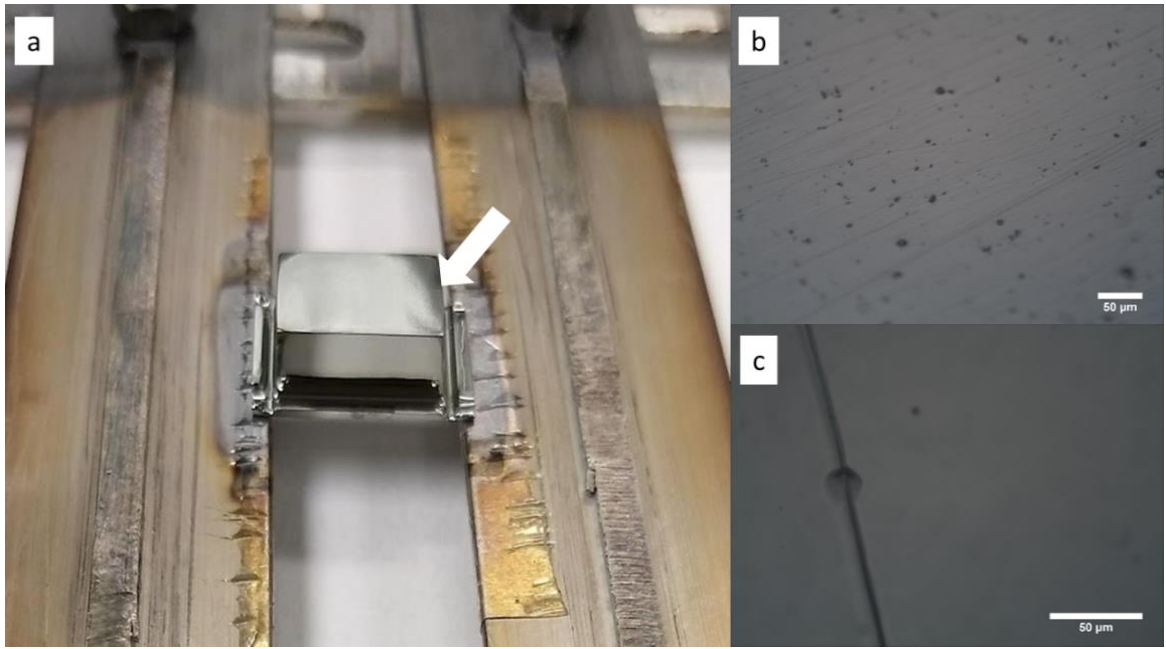

**Figure 11.** (a) A deficient area on the crystal of USD-L02 detector after chemical etching (white arrow pointed to). (b) Microscopic image around the deficient area after manually lapping. (c) Linear defect appeared after the chemical etching. Scale bar 50 µm.

In addition, one must thoroughly remove wax from the entire well-cut crystal. Any invisible wax left on the crystal may cause a surface defect during long-term etching because such residual wax can block chemical etching on the covered area. Extra attention must be given to the four groove area where wax is very likely to stick.

### 3.3 Detector Characterization

All eight detectors were measured for their leakage current at liquid nitrogen temperature to determine the property of the a-Ge contact. Their C-V characteristic was also measured to obtain the full depletion voltage for the calculation of impurity concentration of crystal by using the

– 12 –

equation $N_{|A-D|} = 2\varepsilon_{Ge}\varepsilon_0 v_{fd}/ed^2$, as described earlier, where $v_{fd}$ is the fully depleted voltage. The details of detector characteristics, study of contact property, and related calculations are published in another paper from our group [71]. The current paper will not discuss how we characterize the detector and convert the experimental data in detail. The following part will focus on the USD-L07 detector as an example of detector characterization.

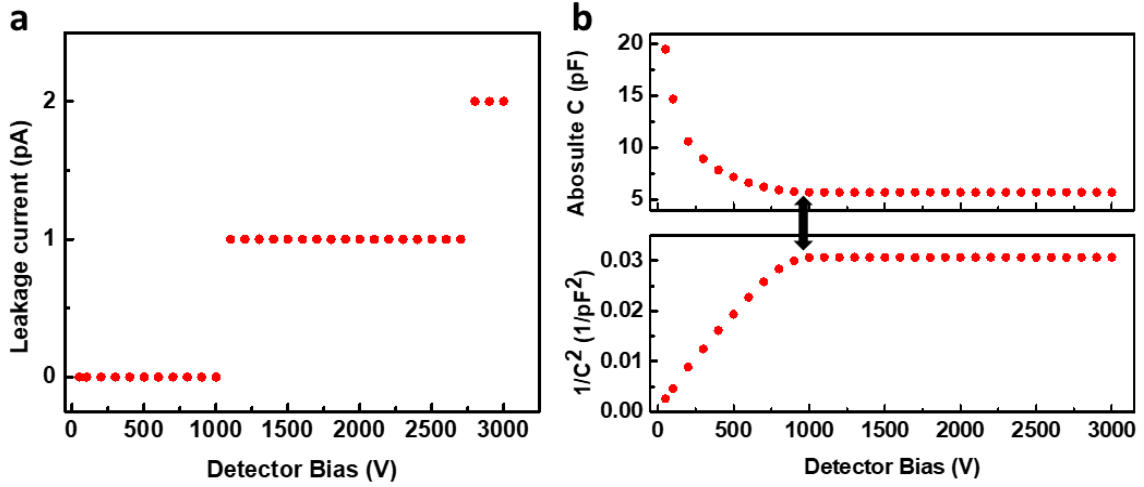

**Figure 12.** (a) Leakage current as function of applied bias measurement for detector of USD-L07. (b) Plots of C-$V_{ap}$ and $1/C^2$ vs. $V_{ap}$.

Figure 12a illustrates the leakage current measurement of USD-L07 at 79 K from the third thermocycle test, which was taken 4 months after its fabrication. The incremental steps in leakage current is caused by the limitation of our measuring instrument. The leakage current was still very low at about 2 pA when the applied voltage was increased up to 3000 V. This means that both a-Ge contacts and Al layers are very suitable for long-term use and an a-Ge contact can effectively block both holes and electrons. Figure 12b shows the absolute capacitance of the detector at the corresponding applied voltage (C-$V_{ap}$) and the plot of $1/C^2$ vs. the bias voltage. Both plots contributed to the determination of full depletion voltage since the absolute capacitance should be a constant once the detector is fully depleted. From both plots, the full depletion voltage can be determined at around 1000 V. Then the impurity concentration of the crystal was calculated through the equation above and found to have a value of $1.97 \times 10^{10}$ cm$^{-3}$, which is an averaged impurity conventration across the entire crystal. Note that the impurity concentration determiend this way is more accurate than that of the Hall Effect measurements. This information was fed back to our crystal-growth group for the improvement of crystal quality.

Energy resolution plays an important role in judging detector performance. We used USD-L07 to collect an energy spectrum of a Cs-137 source with a radioactivity of 5.0 μCi. The Cs-137 source was put on the top of the cryostat right above the top surface of the detector. Negative voltage of 1500 V was applied to the bottom of the detector. Data collection took one hour. The energy spectrum obtained at 79 K is shown in Figure 13. A pulser peak displayed the electronic noise of the test system. The full width of half maximum (FWHM) at 662 keV was 1.62 keV. The pulser shows a FWHM of 1.19 keV to represent the noise level. The energy resolution at 662 keV was 0.197%, which is very close to that reported of commercial detectors (0.20% at 662 keV) [72-73]. All other fully depleted four-wing detectors in Table 1 displayed very similar energy resolution.

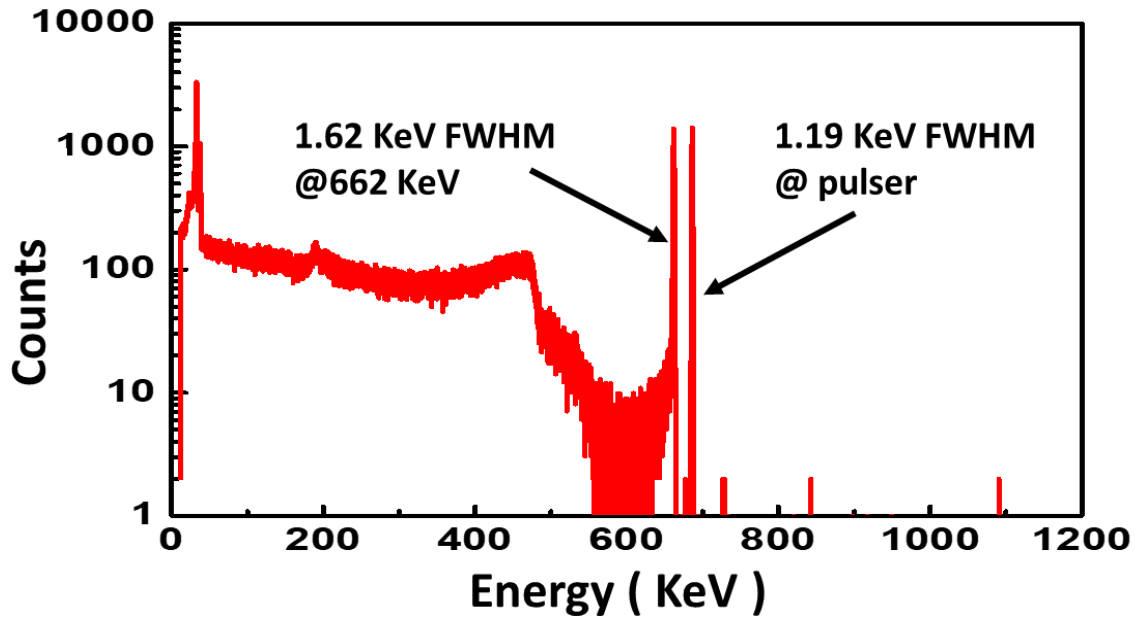

**Figure 13.** Energy spectrum of Cs-137 source collected through the USD-L07 detector. A special condition set was applied voltage -1500 V and the data collecting time was 1 hour.

## 4. Conclusions

The fabrication process of small planar HPGe detectors has been presented in detail beginning with the cutting of a high quality Ge crystal grown at USD. Each step, cutting, lapping, and chemical etching is very critical and directly determines the detector performance. Note that a slow feeding speed in the cutting process will help to avoid mechanical damage to the crystal. A uniform surface texture can be obtained after manually lapping with two different size abrasives. Shiny mirror-like surfaces can be achieved through a long-term chemical etching process. A small cloudy area may not critically impact the detector performance, but any obvious cracks and severe scratches must be removed if they appear after chemical etching. A well-known a-Ge semiconductor technique was employed to passivate the side surfaces of the detector and to form the contact layers on the top and bottom surfaces to block both electrons and holes. The thin Al layers were coated on the a-Ge contacts for signal readout. Such a-Ge semiconductor technology is a simple and efficient method to fabricate HPGe detectors at USD.

Eight planar HPGe detectors have been fabricated at USD over six months. Four detectors (USD-L01, USD-L06, USD-L07, and USD-L08) displayed very good performance with low leakage current and excellent energy resolution for spectroscopic measurement at the temperature of 79 K. USD-L02 was found to have reported defects and it did not work as a detector. USD-L03, USD-L05 and USD-L06 were made with two wings and only USD-L06 worked as a good Ge detector. This indicates that the detectors with two wings can suffer high leakage current resulting from the fabrication and handling processes, which caused a failure in the performance of detectors. Therefore, it is highly prerferable to make detectors with four wings. USD-L04 was successfully fabricated into a detector. However, it cannot be fully depleted at 3700 volts due to its thickness of 10.7 mm, which requires a full depletion voltage of ~4000 volts for a given impurity level of ~$4\times10^{10}$/cm$^3$. This reveals the constraints of the detector thickness and its impurity level for a planar detector.

In summary, we have shown that the fabrication of a good Ge detector does not only require a good quality of Ge crystal, but also a reliable fabrication process. For the former, the Ge crystals must meet the requirement of the impurity level defined by the detector geometry and must be free of linear defects as illustrated with USD-L02. For the latter, the detector fabrication and the detector handling processes are critical. Therefore, the detectors with four wings have a higher success rate than that of two wings. These results demonstrate that USD can not only grow high quality detector-grade germanium crystals of variable size, but also is capable of successfully fabricating detectors with acceptable performance based on measured impurity levels. In addition, guard-ring planar detectors and P-type point contact Ge detectors are currently under investigation by our group. A large cryostat was designed at USD and is currently being constructed for characterization of enlarged planar detectors in the near future.

**Acknowledgments**

The authors would like to thank Dr. Mark Amman for his supervision on developing Ge planar detectors with a-Ge contacts at USD and Dr. Christina Keller for a careful reading of the manuscript. We would also like to thank the Nuclear Science Division at Lawrence Berkeley National Laboratory for providing us a testing cryostat. This work was supported in part by NSF OISE 1743790, NSF OIA-1738695, NSF OIA-1738632, DOE grant DE-SC0004768 (DE-FG02-10ER46709), the South Dakota Board of Regents Innovation Grant, the Office of Research at the University of South Dakota and a research center supported by the State of South Dakota.